\documentclass{emulateapj}
\usepackage{amsmath}
\usepackage{amssymb}
\usepackage{graphicx}
\usepackage{appendix}
\usepackage{color}
\usepackage{multirow}
\usepackage{rotating}
\newcommand{\lSect}[1]{{\label{sec:#1}}}
\newcommand{\lFig}[1]{{\label{fig:#1}}}
\newcommand{\lEq}[1]{{\label{eq:#1}}}
\newcommand{\lTab}[1]{{\label{tab:#1}}}
\newcommand{\FIGFF}[2]{{\ref{fig:#2}{#1}}}

\newcommand{\FIG}[2]{{Fig.~\FIGFF{#1}{#2}}}
\newcommand{\Fig}[1]{{\FIG{}{#1}}}

\newcommand{\Sectff}[1]{{\ref{sec:#1}}}
\newcommand{\Sect}[1]{{\S~\Sectff{#1}}}

\newcommand{\Eqref}[1]{{\ref{eq:#1}}}
\newcommand{\Eqff}[1]{{(\Eqref{#1})}}

\newcommand{\Eq}[1]{{eq.~\Eqff{#1}}}

\newcommand{\Msun}{\ensuremath{\mathrm{M}_\odot}}
\newcommand{\Rsun}{\ensuremath{\mathrm{R}_\odot}}

\newcommand{\Tab}[1]{{Table \ref{tab:#1}}}

\def\gtaprx {\lower .1ex\hbox{\rlap{\raise .6ex\hbox{\hskip .3ex
	{\ifmmode{\scriptscriptstyle >}\else
		{$\scriptscriptstyle >$}\fi}}}
	\kern -.4ex{\ifmmode{\scriptscriptstyle \sim}\else
		{$\scriptscriptstyle\sim$}\fi}}}
\def\ltaprx {\lower .1ex\hbox{\rlap{\raise .6ex\hbox{\hskip .3ex
	{\ifmmode{\scriptscriptstyle <}\else
		{$\scriptscriptstyle <$}\fi}}}
	\kern -.4ex{\ifmmode{\scriptscriptstyle \sim}\else
		{$\scriptscriptstyle\sim$}\fi}}}

\begin{document}

\submitted{15 Feb, 2016}
\accepted{-- ---, 2016}

\title{The Most Luminous Supernovae}

\author{Tuguldur Sukhbold\altaffilmark{1} and 
             S.\ E.\ Woosley\altaffilmark{1}}
\altaffiltext{1}{Department of Astronomy and Astrophysics, 
                         University of California, Santa Cruz, CA 
                         95064. sukhbold@ucolick.org}

\begin{abstract}
Recent observations have revealed a stunning diversity of extremely
luminous supernovae, seemingly increasing in radiant energy without
bound. We consider simple approximate limits for what existing
models can provide for the peak luminosity and total radiated energy
for non-relativistic, isotropic stellar explosions. The brightest
possible supernova is a Type I explosion powered by a sub-millisecond
magnetar with filed strength $B \sim$ few $\times 10^{13}$ G. In
  extreme cases, such models might reach a peak luminosity of
$\rm 2\times10^{46}\ erg\ s^{-1}$ and radiate a total energy of up to
$\rm 4\times10^{52}\ erg$. Other less luminous models are also
explored, including prompt hyper-energetic explosions in red
supergiants, pulsational-pair instability supernovae, pair-instability
supernovae and colliding shells. Approximate analytic
expressions and limits are given for each case. Excluding
  magnetars, the peak luminosity is near $\rm 3\times
  10^{44}\ erg\ s^{-1}$ for the brightest models and the corresponding
  limit on total radiated energy is $\rm 3\times
  10^{51}\ erg$. Barring new physics, supernovae with a light output
  over $3 \times 10^{51}$ erg must be rotationally powered, either
  during the explosion itself or after, the most obvious candidate
  being a rapidly rotating magnetar.  A magnetar-based model for the
recent transient event, ASASSN-15lh is presented that strains, but
does not exceed the limits of what the model can provide.
\end{abstract}


\section{Introduction}
\lSect{intro}

Motivated by the recent discovery of many ultra-luminous supernovae 
(ULSN), including, controversially, the extreme case of 
ASASSN-15lh \citep{Don16,Bro15}, the limits of several scenarios often 
invoked for their interpretation are considered. These include colliding 
shells, pair-instability supernovae, and newly-born magnetars 
\citep[e.g.][]{Gal12,Qui11}. Each of these energy sources will give 
different results when occurring in a stripped core of helium or 
carbon and oxygen (Type I) or a supergiant (Type II), and both 
cases are considered.  All calculations of explosions and light curves 
use the 1D implicit hydrodynamics code \texttt{KEPLER} 
\citep{Wea78,Woo02}, and employ presupernova models that have 
been published previously.

The more extreme case of ``relativistic supernovae'' - either 
supernovae with relativistic jets or the explosion of super-massive 
stars that collapse because of general relativistic instability
\citep{Ful86,Che14} is not considered here. These are rare events 
with their own distinguishing characteristics.

\section{Prompt Explosions and Pair-Instability}
\lSect{prompt}

Any explosive energy that deposits before the ejecta significantly
expands will suffer severe adiabatic degradation that will prevent the
supernova from being particularly bright. An upper bound for prompt
energy deposition in a purely neutrino-powered explosion is $\rm
\sim2-3\times10^{51}\ erg$ \citep{Fry01,Ugl12,Pej15,Ert16}, which is
capable of explaining common supernovae \citep{Suk15}, but not the
more luminous ones.  In a red, or worse, blue supergiant, the
expansion from an initial stellar radius of, at most, 10$^{14}$ cm, to
a few times $10^{15}$ cm, where recombination occurs, degrades the
total electromagnetic energy available to $\rm \ltaprx
10^{50}\ erg$. Even in the most extreme hypothetical case, where a
substantial fraction of a neutron star binding energy, $\sim10^{53}$
erg, deposits instantly, the light curve is limited to a peak
brightness of approximately $\rm 10^{44}\ erg\ s^{-1}$ (neglecting the
very brief phase of shock break out).

This can be demonstrated analytically and numerically. Adopting the
expression for plateau luminosity and duration from \citet{Pop93} and
\citet{Kas09}, as calibrated to numerical models by \citet{Suk15},
Type II supernovae have a luminosity on their plateaus of
\begin{equation}
\rm L_{p} = 8.5 \times10^{43} \ R_{0,500}^{2/3} M_{env,10}^{-1/2} E_{53}^{5/6}\ erg\ s^{-1},
\lEq{L25}
\end{equation}
where $\rm R_{0,500}$ is the progenitor radius in 500 \Rsun, $\rm
M_{env,10}$ is the envelope mass in 10 \Msun, and $\rm E_{53}
  \ltaprx 1$ is the prompt explosion energy in units of 10$^{53}$
erg. The approximate duration of the plateau, ignoring the effects of
radioactivity, is given by
\begin{equation}
\rm \tau_p = 41 \ E_{53}^{-1/6} M_{10}^{1/2} R_{0,500}^{1/6} \ days.
\lEq{tauplat}
\end{equation} 
This plateau duration is significantly shorter than common 
supernovae due to the much higher energies considered.

These relations compare favorably with a model for a 15
\Msun\ explosion calculated with an assumed explosion energy of $\rm
0.5\times10^{53}\ erg$ (\Fig{rsg_pist}). Here the red supergiant
presupernova stellar model from \citet{Woo07} had a radius of 830
\Rsun \ and an envelope mass, 8.5 \Msun. The estimated luminosity on
the plateau from \Eq{L25} is $\rm 6.7\times 10^{43}\ erg\ s^{-1}$ and
duration from \Eq{tauplat} is 46 days. The corresponding \texttt{KEPLER} model
in \Fig{rsg_pist} had a duration of $\sim45$ days and a luminosity at
day 25 of $\rm 6.6\times 10^{43}\ erg\ s^{-1}$. The total energy
emitted is approximately $\rm L_p \tau_p$, or $\rm 3 \times 10^{50}
E_{53}^{2/3} R_{0,500}^{5/6}$ erg.

\begin{figure}[h]
\centering \includegraphics[width=.48\textwidth]{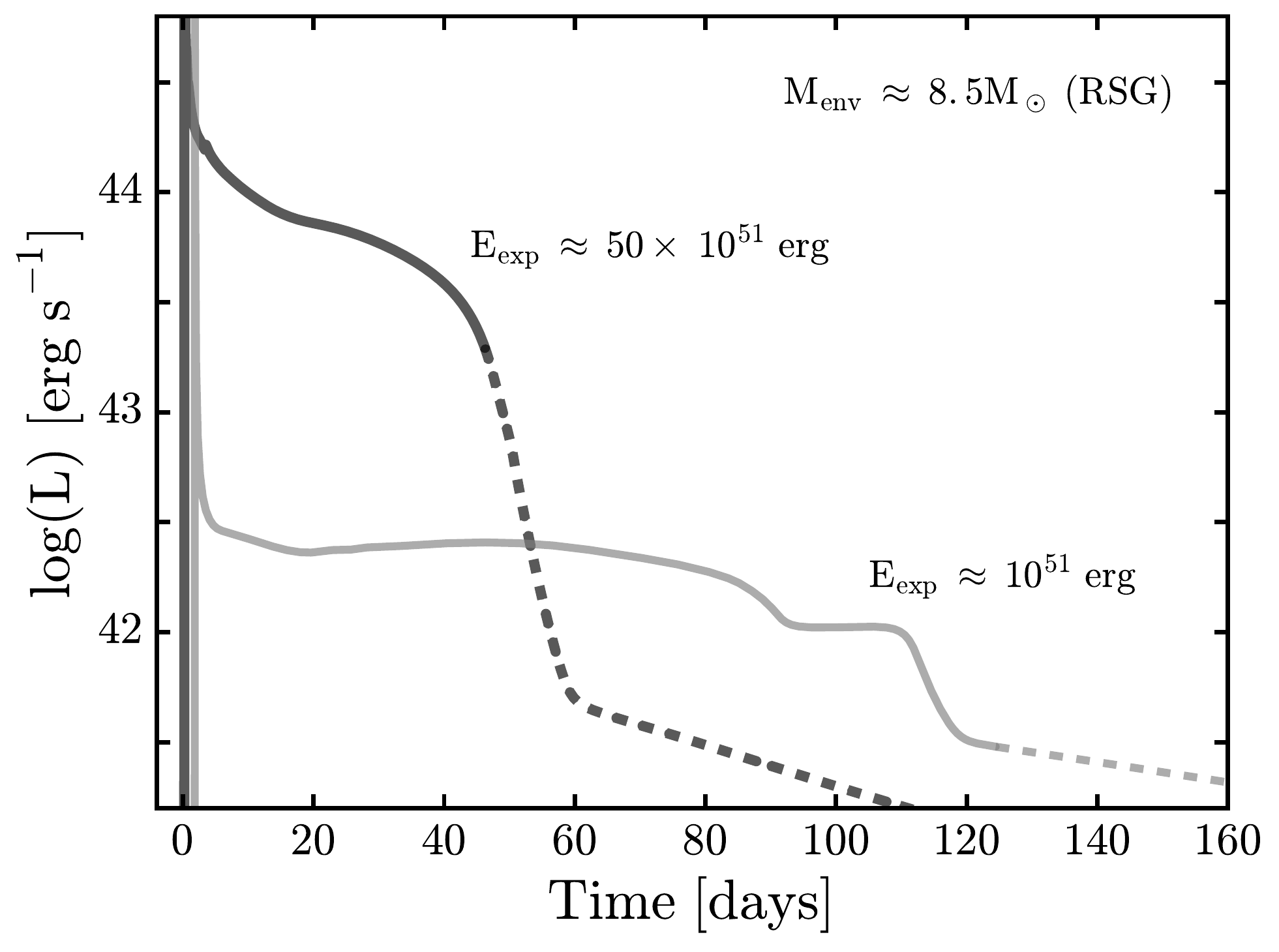}
\caption{Bolometric light curves for a 15 \Msun\ supergiant exploded
  with two different values of prompt energy deposition.  One with 
  $\rm E_{exp}=10^{51}$ erg, is typical of common Type IIp 
  supernovae; the other with $\rm E_{exp}=50 \times 10^{51}$ erg is 
  near the upper bound of what any prompt,
  point explosion might provide. Even for this extreme case, the
  plateau luminosity does not exceed $\rm
  \sim10^{44}\ erg\ s^{-1}$. The curves are dashed when the ejecta
  become optically thin and the blackbody representation of their
  emission becomes questionable. The presupernova star, originally 15
  \Msun\ at birth, had a mass of 12.6 \Msun, of which 8.5 \Msun\ was
  in the hydrogen envelope, and a radius of $\rm R_{0,500}\sim1.7$. The
  luminosity at shock break out in the more energetic model peaked at $\rm
  1.5 \times 10^{47}\ erg\ s^{-1}$, but only lasted for about 100
  seconds.  \lFig{rsg_pist}}
\end{figure}

Similar limits apply to pair-instability supernovae (PISN) in 
red supergiant progenitors.  Again, maximum explosion energies are
$\ltaprx 10^{53}$ erg \citep{Heg02}. For the most extreme, rarest
case, $\rm M_{10}\approx 20$, $\rm R_{0,500} \approx 5$ and $\rm
E_{53} \approx 1$, \Eq{L25} and \Eq{tauplat} imply plateau
luminosities near $\rm 5\times10^{43}\ erg\ s^{-1}$ for about 200
days.  These values are consistent with the \texttt{KEPLER} models given in
\citet{Sca05}. The total radiated energy is $\rm 1 \times 10^{51}$
erg. Most of the radioactivity decays during the plateau. Since the decay 
energy is substantially less than the explosion energy, the
modification of the light curve during its bright plateau is not
appreciable.

We conclude that ULSN must be energized by a power source that
  deposits its energy well after the original explosion.  The known
  delayed energy sources are radioactivity, colliding winds, and
  pulsars.

\section{Radioactivity}

The most prolific sources of $^{56}$Ni are PISN.  The rarest, most 
massive PISN produces, at most, 50
\Msun\ of $^{56}$Ni in an explosion with a final kinetic energy of $\rm
9\times 10^{52}\ erg$ \citep{Heg02}.  This large production only
occurs for the the most massive helium cores ($\sim$130 \Msun), which
very nearly collapse to black holes.  The total energy available from
the decay of large amount of $^{56}$Ni is substantial,
\begin{equation}
\rm E_{dec}\ \approx\ 2.4\times10^{51}\Big(\frac{M_{Ni}}{50\Msun}\Big)(3e^{-t/\tau_{Co}}+e^{-t/\tau_{Ni}})\ erg,
\end{equation}
where $\rm M_{Ni}$ is the $^{56}$Ni mass in \Msun, and $\rm
\tau_{Co}=111$ d and $\rm \tau_{Ni}=8.7$ d are mean lives of $^{56}$Co
and $^{56}$Ni. For a nickel mass of $\sim$50 \Msun\ the total energy
is nearly $10^{52}$ erg. Most of this energy is lost during the
adiabatic expansion to peak, however.

For a star that has lost its hydrogen envelope, an approximate
estimate when the PISN light curve peaks is given by equating the
effective diffusion timescale, $\rm t_d$, to age. This gives a time of
peak luminosity, $\rm t_p$, of
\begin{equation}
\begin{aligned}
\rm t_p&=\rm \Big(\frac{3\kappa}{4\pi c}\Big)^{1/2}\ \Big(\frac{M_{ej}^3}{2E_{exp}}\Big)^{1/4}\\
 &\rm \sim177\ \Big(\frac{M_{ej}}{130\Msun}\Big)^{3/4}\Big(\frac{E_{exp}}{10^{53}erg}\Big)^{-1/4}\ days,
\end{aligned}
\end{equation}
where $\rm M_{ej}$ is the ejecta mass in \Msun\ and $\rm E_{exp}$ 
is the explosion energy in erg. Considering the similarity of high 
velocity and iron-rich composition to Type Ia supernovae, an opacity
$\rm \kappa\approx \rm 0.1\ cm^2\ g^{-1}$ is assumed. Arnett's 
Rule \citep{Arn79} then implies a maximum luminosity of
\begin{equation}
\rm L_p\approx 8\times10^{44}\
\Big(\frac{M_{Ni}}{50\Msun}\Big)e^{-t_p/\tau_{Co}}\
erg\ s^{-1}.
\end{equation}
Only the luminosity due to the decay of $^{56}$Co is included here, 
since for t $\rm \sim t_p$, most of $^{56}$Ni will have already 
decayed. For the fiducial values of $\rm t_p$ and $\rm M_{Ni}$, the 
peak luminosity is then roughly $\rm 1.5\times10^{44}\ erg\ s^{-1}$, 
which compares favourably with models in which the hydrodynamics 
and radiation transport are treated carefully \citep{Kas11,Koz15}
and with the analytic models of \citep{Cha13}.

Assuming that the total emitted energy by a Type I supernova is 
\begin{equation}
\rm E_{rad}\approx\frac{1}{2}L_pt_p+E_{dec}(t_p),
\lEq{erad_type1}
\end{equation}
and using the fiducial values, the approximate upper bound on the
total luminous energy in a PISN-Type I is $2.6\times10^{51}$ erg, about
one quarter of the total decay energy.


\section{Colliding Shells}
\lSect{shells}

\subsection{Generic Models}
\lSect{generic}

Observations show that a substantial fraction of ULSN, especially
those of Type IIn, are brightened by circumstellar interaction
\citep[e.g.][]{Kie12}. In some cases this interaction can be extremely
luminous \citep{Smi10,Smi11a}. The necessary mass loss is often
attributed to prior outbursts of the star as a luminous blue variable
\citep[LBV; e.g.][]{Smi11b} or a pulsational-pair-instability supernova
\citep[PPISN,][]{Woo07}, though other possibilities, e.g. common
envelope \citep{Che12a}, are sometimes invoked.

The luminosity of colliding shells is limited by their differential
speeds, their masses, and the radii at which they collide. If the
collision happens at too small a radius where the ejecta is still very
optically thick, colliding shells become another variant of ``prompt
explosions'' (\Sect{prompt}). On the other hand, if the collision
happens at too large a radius, the resulting transient has a longer
time scale, lower luminosity, and may not emit chiefly in the optical
\citep{Che12b}. In practice, these constraints limit the radius where
the shells collide and produce a bright optical transient to roughly
$10^{15}$ - 10$^{16}$ cm.  A similar range of radii is obtained by
multiplying typical collision speeds, $\sim5,000$ km s$^{-1}$,
  by the duration of an ULSN, $\sim 100$ days.

\citet{Che12b} give a maximal ``cooling luminosity'' for
  colliding shells in which most of the dissipated energy goes into
  light \citep[see also][eq. 1]{Smi10},
\begin{equation}
\begin{split}
\rm L \ & \ltaprx \rm \ 2 \pi r^2 \rho v_{shock}^3=0.5 \frac{\dot M}{v_{wind}} v_{shock}^3 
\end{split}
\end{equation}
where $\rm \dot M$ is the pre-explosive mass loss rate with speed 
$\rm v_{wind}$, and $\rm v_{shock}$ is the shock speed of the explosive
ejecta impacting that ``wind''. Narrow lines in the spectra of Type
 IIn supernovae, including some very luminous ones \citep{Kie12},
imply pre-explosion wind speeds of a few hundred to 1000 km
s$^{-1}$. At those speeds, and given that the light curve is generated
at $\rm r \sim 10^{15} - 10^{16}\ cm$, the relevant time for the mass loss
is a few years before the final explosion. The velocity of the shock
is $\rm v_{shock} \approx \sqrt{2 E/M}$, where E is the explosion
energy of mass M. Here we normalize it to 10$^9$ cm s$^{-1}$ as in
\citet{Che12b}, though it implies a very energetic explosion. The luminosity 
from the collision is then
\begin{equation}
\rm L \approx 3.1\times 10^{44} \frac{\dot M_{-1}}{v_{wind,7}}
v_{shock,9}^3 \ erg \ s^{-1},
\end{equation}
where $\rm \dot M_{-1}$ is the mass loss rate a few years before the
explosion normalized to 0.1 \Msun \ per year, $\rm v_{wind,7}$ is in
$\rm 10^2\ km\ s^{-1}$ and $\rm v_{shock,9}$ is in units of $\rm
10^4\ km\ s^{-1}$. Typical values for the outbursts that produce very
bright bright Type IIn supernovae are $\rm \dot M_{-1}$ = 1, $\rm
v_{wind,7}$ = 1 to 10, and $\rm v_{shock,9}$ = 0.5 \citep{Kie12}
implying peak luminosities near $5 \times 10^{43}$ erg s$^{-1}$. The
large ejection in $\eta$-Carina in the 1840's ejected 12 \Msun
\ moving at $v_{\rm wind,7}$ up to 6.5 \citep{Smi08}.

It is the mass of the shell into which a supernova of given energy
plows that matters most \citep{Van10}. We are unaware of any models
other than PPISN (\Sect{ppsn} or 10 \Msun \ stars \citep{Woo15a} that
eject solar masses of material just years before dying. If a generous
upper limit of 10 \Msun \ between 10$^{15}$ and 10$^{16}$ cm ($M_{10}$
= 1) is adopted, $\rm \dot M_{-1}/v_{wind,7}$ is $\rm \ltaprx \, 3 \,
M_{10} \, R_{16}^{-1}$ where $\rm R_{16}$ is the outer edge of the
interaction region in 10$^{16}$ cm units. For a shock speed $\rm
v_{shock,9} = 0.5$ and an event duration of 100 days, $\rm R_{16} \sim
0.5$. The maximum luminosity is then $\rm 10^{45} v_{shock,9}^3
M_{10}R_{16}^{-1}\ erg\ s^{-1}\approx 3 \times 10^{44}\ erg\ s^{-1}$.
More generally the maximum luminosity is $\rm L = 10^{45} \tau_{\rm
  SN,100}^{-1} M_{10} v_{\rm shock,9}^2\ erg\ s^{-1}$ where $\tau_{\rm
  SN,100}$ is the duration of the brightest part of the light curve.
This limit is sufficient to accommodate all ULSN that maybe powered by
collisions and is consistent with the theoretical results of
\citet{Van10}.

It might be possible to raise this limit by invoking slightly greater
shock speeds or shell masses. The former requires extremely energetic
supernovae though. Accelerating a shell of 10 \Msun \ to 10$^9$ cm
s$^{-1}$ requires an explosion energy of at least 10$^{52}$ erg and
100\% conversion efficiency. This is considerably more than neutrinos
can provide and already indicates a source that is, at heart,
rotationally powered. Yet it may be that having high mass loss rates
removes sufficient angular momentum to inhibit the formation of
rapidly rotating iron cores. Even energetic PISN do not develop speeds
of 10,000 km s$^{-1}$ in a significant part of their mass. Moreover,
PISN are burning carbon radiatively in their centers the last few
years of their life and, except for PPISN (\Sect{ppsn}), experience no
obvious instability that would lead to the impulsive ejection of 10
\Msun.

With considerable uncertainty, we thus adopt an upper limit for
colliding shells of $\rm 3\times 10^{44}\ erg\ s^{-1}$ and a total
radiated energy of $\rm \tau_{SN} L \sim 3 \times 10^{51}$ erg.  For
bare helium cores which are not PPISN and clearly not LBVs, the values
are likely to be much smaller because of the smaller shell masses, but
existing models, do not allow a specific estimate.

\subsection{Pulsational-pair instability supernovae}
\lSect{ppsn}

The most luminous colliding shell models with definite predictions for
their luminosity are PPISN \citep{Woo07,Yos16}.  For a narrow range
of masses corresponding to stars with 50 - 55 \Msun\ helium cores, a
supergiant star, red or blue, will eject its hydrogen envelope at
speeds $\sim1000$ km s$^{-1}$, and a year or so later eject one or
more very energetic shells that smash into it \citep{Woo15b}. The
source of the energy is the thermonuclear burning of carbon and
oxygen. For lighter helium cores, low energy shells are ejected in
rapid succession before the envelope has expanded to 10$^{15}$ cm. The
collision energy is adiabatically degraded and the resulting supernova
is not especially luminous (Woosley 2016, in prep.). For heavier
cores, the pulses are too infrequent and produce collisions outside of
10$^{16}$ cm that last much longer than one hundred days.

In the narrow helium-core mass range of 50 - 55 \Msun\ though, one or
more pulses occurring a year or so after the one that ejects the
envelope, eject additional shells carrying a energy of up to $1 \times
10^{51}$ erg \citep{Woo15b}. Radiating all this energy over a 10$^{7}$
s interval gives a luminosity that can approach 10$^{44}$ erg s$^{-1}$
\citep{Woo07}.

For helium cores lacking any hydrogen envelope the luminosities are
less because of the lack of a massive low velocity reservoir to turn
kinetic energy into light. Typical peak luminosities for Type I PPISN
are thus near $3 \times 10^{43}$ erg s$^{-1}$, and the light curve can
be more highly structured \citep{Woo15b}.


\section{Magnetars}
\lSect{magnetars}

With some tuning, the energy deposited by a young magnetar in the
ejecta of a supernova can significantly brighten its light curve
\citep{Mae07,Woo10,Kas10}. The model has been successfully applied to
numerous observations of Type Ic ULSN \citep[e.g.][]{Ins13,Nic13,How13} 
and magnetars seem a natural consequence of collapse of rapidly rotating 
cores \citep[e.g.][]{Mos15}.

The rotational kinetic energy of a magnetar with a period $\rm
P_{ms}=P/ms$ is approximately $\rm
E_m\approx2\times10^{52}P_{ms}^{-2}$ erg, where $\rm
E_{m,max}\approx4\times10^{52}$ erg is the rotational energy for an
initial period of $\sim$0.7 ms. Usually this period is
  restricted to $> 1$ms, because of rotational instabilities that lead
  to copious gravitational radiation. However, \citet{Met15} have
  recently discussed the possibility that the limiting rotational
  kinetic energy could exceed $10^{53}$ erg, depending on the neutron
  star mass and the equation of state and here we adopt that value as
  an upper bound. This energy reservoir can be tapped through vacuum
dipole emission, which is approximately $\rm
E_m/t_m\approx10^{49}B_{15}^2P_{ms}^{-4}\ erg\ s^{-1}$, where $\rm
B_{15}=B/10^{15}$ G is the dipole field strength at the equator, and
$\rm t_m=2\times10^3P_{ms}^2B_{15}^{-2}$ s is the magnetar spin-down
timescale. A magnetic dipole moment B($10$km)$^3$ is adopted, and an
angle of $\pi/6$ between the magnetic and rotational axes has been
assumed. Combining these relations, one obtains the temporal evolution
of the rotational energy and magnetar luminosity as $\rm
E_m(t)=E_{m,0}t_m/(t_m+t)$ and $\rm L_m(t)=E_{m,0}t_m/(t_m+t)^2$.

The peak luminosity can be estimated using the diffusion equation and 
ignoring the radiative losses in the first law of thermodynamics 
\citep{Kas10}:
\begin{equation}
\rm L_{p}=\rm \frac{E_{m,0}}{t_d}\bigg[\rm \xi\ ln\big(1+\frac{1}{\xi}\big)-\frac{\xi}{1+\xi} \bigg],
\lEq{lpeak}
\end{equation}
where $\rm \xi=t_m/t_d$ is the ratio of spin-down to effective 
diffusion timescales. The term inside square brackets has a maximum 
at $\rm \xi\approx1/2$, obtained by solving 
$\rm d(L_pt_d/E_{m,0})/d\xi=0$. This implies an optimal field 
strength for maximizing the peak luminosity is:
\begin{equation}
\rm B_{15}\Big|_{L_{p,max}}\simeq 66P_{ms,0} t_d^{-1/2}.
\lEq{bp_peak}
\end{equation}
That is, for a given combination of $\rm P_{ms,0}$ and ejecta
parameters - $\rm M_{ej}, E_{sn}, \kappa$, the brightest possible 
peak luminosity is obtained for this field strength. 

The maximum peak luminosity is then $\rm L_{p,max}\simeq
E_m/10t_m$. For the limiting initial spin of $\rm P_{ms,0}=0.7ms$ the corresponding field strength is $\rm B\approx4\times10^{13}$ 
G (for $\rm \kappa=0.1\ cm^2\ g^{-1}$, $\rm M_{ej}=3.5$ \Msun\ and $\rm 
E_{SN}=1.2\times10^{51}\ erg$), and the limiting peak luminosity is 
$\rm L_{p,max}\approx2\times10^{46}\ erg\ s^{-1}$. Here 
$\rm E_m\gg E_{SN}$, therefore unless one invokes even lower 
$\rm \kappa$, $\rm M_{ej}$ and much larger $\rm E_m$, any 
transient with brighter observed luminosity will be hard to explain 
by the magnetar model (\Fig{lcs}).

Using a version of Arnett's Rule \citep[e.g.]{Ins13}, $\rm
L_m(t_p)=L_p$, the time for $\rm L_p$ is $\rm t_p = (E_m t_m
L_p^{-1})^{1/2}-t_m$. For the maximal luminosity, the corresponding
peak time is then $\rm t_{p,max}\simeq2.2 t_m$. This can be used to
estimate the limiting radiated energy in the same way as in
\Eq{erad_type1} to find that:
\begin{equation}
\rm E_{rad,max}\simeq0.4E_{m,0}.
\end{equation}
Any observation with a total radiated energy of $\rm E_{rad}>4\times10^{52}$
erg will be nearly impossible to explain by the magnetar model. A more
  conventional value and one that fits ASASSN-151h (\Fig{lcs}), is $2
  \times 10^{52}$ erg. This is within a factor of two of the limiting
  magnetar kinetic energy inferred for gamma-ray bursts by \citet{Maz14}.

\begin{figure}[h]
\centering
\includegraphics[width=.48\textwidth]{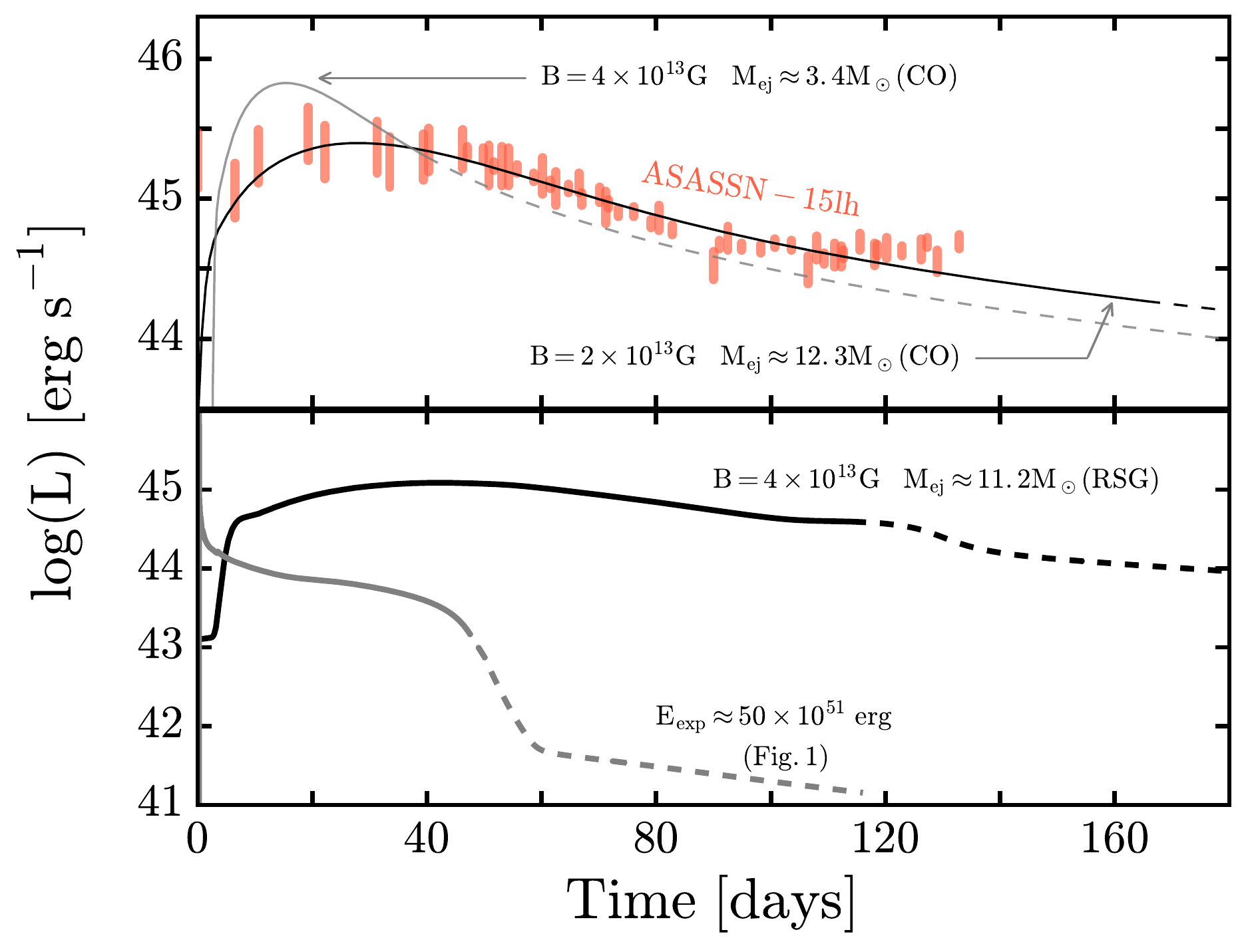}
\caption{Top: The luminous transient ASASSN-15lh compared with a
  magnetar model {\rm in which the initial rotational energy was $5
    \times 10^{52}$ erg}. Similar magnetars embedded in a less ejecta
  will give slightly brighter light curves.  Bottom: The same magnetar
  used in the top panel for fitting ASASSN-15lh is embedded in the
  ejecta of a massive red supergiant progenitor. The light curve is
  dimmer than the Type I case, but substantially brighter than the
  prompt explosion case shown in \Fig{rsg_pist}. The dashed curve
  marks the transition to nebular phase.  \lFig{lcs}}
\end{figure}

To illustrate these limits, a series of magnetar-powered models based
upon exploding CO cores \citep[from][]{Suk14} was calculated to find a
best fit to the light curve of ASASSN-15lh. In each case, soon after
bounce, the magnetar deposited its energy in the inner ejecta at a
rate given by the vacuum dipole spin-down rate. The top panel of
\Fig{lcs} shows the best fitting model, which employs a magnetar with
an initial period of 0.7 ms and magnetic field strength of $\rm
2\times10^{13}$ G, illuminating the ejecta in the explosion of a 14
\Msun\ CO core ($\rm M_{ej}\approx11.2\Msun$).

These magnetar parameters agree well with the previously published
fits, but the ejecta masses are different. An ejecta mass of 15
\Msun\ was obtained by \cite{Dai16}, as they used simple
semi-analytical models, which ignored all dynamical effects and
deviate from hydrodynamic calculations most when $\rm E_{m}\gg
E_{sn}$. An ejecta mass of only 3 \Msun\ was used in \cite{Met15}, as
they applied the same simple semi-analytical model for the early
release of the data spanning only $\sim$60 days.  That fit would not
work for the later data shown in \Fig{lcs}, and the ejecta would
become optically thin at an early time. \cite{Ber16} limited their
models to small He-cores (8 \Msun), and their model does not fit the
broad peak of ASASSN-15lh well.


Magnetars can also illuminate bright, long lasting Type II supernovae.
The bottom panel of \Fig{lcs}, shows the same magnetar that was
applied for the fit to ASASSN-15lh, now embedded inside the remnant 
of a 15 \Msun\ red supergiant progenitor. Because the ejecta mass is much
larger, the light curve is fainter and much broader. The ejecta stays
optically thick for nearly 4 months. Much as radioactive decay
extends the plateau duration by causing ionization, magnetar-deposited
energy also significantly extends the optically thick period.


\section{Conclusions} 
\lSect{conclude}

\begin{deluxetable}{clcc}
\tablewidth{0pt}
\tablecaption{Limiting Peak Luminosities and Radiated Energies}
\tablehead{\colhead{}                                         &
                   \colhead{model$\quad$}                  &
                   \colhead{$\rm L_p\ [erg\ s^{-1}]$} &
                   \colhead{$\rm E_{rad}\ [erg]$}
                   }
\startdata
\parbox[t]{2mm}{\multirow{4}{*}{\rotatebox[origin=c]{90}{Type II}}}
 & prompt     & $1\times10^{44}$ & $3\times10^{50}$ \\
 & PISN         & $5\times10^{43}$ & $1\times10^{51}$ \\
 & collisions  & $3\times10^{44}$ & $3\times10^{51}$ \\
 & PPISN       & $1\times10^{44}$ & $1\times10^{51}$ \\
 & magnetar & $1\times10^{45}$ & $\ 9\times10^{51}$
\vspace{0.1cm}\\
\hline\vspace{-0.1cm}
&&&\\
\parbox[t]{2mm}{\multirow{3}{*}{\rotatebox[origin=c]{90}{Type I}}}
 & PISN         & $2\times10^{44}$ & $3\times10^{51}$ \\
 & PPISN       & $3\times10^{43}$ & $1\times10^{51}$ \\
 & magnetar & $2\times10^{46}$ & $\ \ 4\times10^{52}$
\vspace{0.1cm}
\lTab{summary}
\end{deluxetable}

\Tab{summary} summarizes the maximum luminosity and total luminous
energy for the models considered. Given the various approximations
made, the numbers are probably accurate to a factor of two in most
cases except for shell ``collisions'' where definitive models
are lacking (\Sect{generic}). In all but the magnetar-powered models,
the peak luminosities are a few times 10$^{44}$ erg s$^{-1}$ and 
peak integrated powers are near 10$^{51}$ erg. This is gratifying since 
most ``superluminous supernovae'' are within those bounds 
\citep[e.g.][]{Nic15}.

For point-like explosions, which includes PISN of Type II, the prompt
energy injection is typically degraded by a factor of $\sim100$ by
adiabatic expansion, so even obtaining 10$^{51}$ erg of light
requires an explosion that strains the limits of both neutron star
binding energy (core-collapse supernovae) and thermonuclear energy
(PISN).  The upper bound for PISN-Type I is also well determined by both
analytic scaling rues and numerical models.


For supernovae whose light comes from colliding shells, the
  constraints are less accurate due to lack of knowledge about the
  masses of the shells involved and the supernova explosion energies
  in cases where large impulsive mass loss occurs just before the star
  dies. The limit in the table assumes shock speeds less than 5000 km
  s$^{-1}$ and shell masses less than 10 \Msun. Estimates for PPISN
  are more precise because the mass of the helium core needed to make
  luminous optical supernovae is highly constrained. In order that
the duration of the pulses be years and not months or centuries, the
helium core mass need to be in the range 50 - 55 \Msun\ and that
restricts the energy of the secondary pulses and supernova.


Magnetars are a special case. The limits come from using a
  simple dipole formula in a situation where it has not been
  observationally tested and assuming what some would regard as a high
  limiting rotational energy for neutron stars. Rotation  can
tap an energy reservoir almost as great as the binding energy of the
neutron star and deposit it over an arbitrarily long time scale -
depending on the choice of magnetic field strength. Thus the optical
efficiency for converting rotational energy to light can be (forced to
be) very high.

It is interesting though that the upper bounds for magnetar-powered
light curves are so high. This implies a possible observable
diagnostic. Supernovae that substantially exceed $\rm 3 \times
  10^{44}\ erg\ s^{-1}$ for an extended period and which have total
luminous powers far above $3 \times 10^{51}$ erg should be
considered strong candidates for containing an embedded
magnetar. Similarly, ``supernovae'' that exceed the generous limits
for magnetar power given in \Tab{summary} may not be supernovae at
all.

ASASSN-15lh \citep{Don16} is an interesting case in this regard. 
\Fig{lcs} shows that it can, barely, be accommodated by a magnetar 
model and \Tab{summary} says it must be a magnetar, if it is a 
supernova \citep{Bro15}.

\section{Acknowledgements}

We thank Alex Heger for his contributions in developing the
\texttt{KEPLER} code and Chris Kochanek, Iair Arcavi, Matt Nicholl and
Todd Thompson for useful comments. This work was supported by NASA
(NNX14AH34G).


\end{document}